# Robust UOWC systems against bubble-induced impairments via transmit/receive diversities

Lian-Kuan Chen (陈亮光)[1], Yingjie Shao (邵英婕)[1*], and Rui Deng (邓锐)[1]

[1]Department of Information Engineering, The Chinese University of Hong Kong, Hong Kong SAR

*Corresponding author: sy017@ie.edu.cuhk.hk

We systematically investigate the bubble-induced performance degradation for underwater wireless optical communication (UOWC) with different bubble sizes and positions. By using different transmit and receive diversities, we investigate the effectiveness of transmit/receive diversity on the mitigation of the bubble-induced impairment to the UOWC link. With the help of a 2×2 MIMO using repetition coding (RC) and maximum ratio combining (MRC), a robust 780-Mbit/s UOWC transmission is achieved. The corresponding outage probability can be significantly reduced from 34.6% for the system without diversity to less than 1%.

*doi:10.3788/COLXXXXXX.XXXXXX.*

To explore the much-unexcavated resources in the ocean, human's underwater activities have been intensified. Leveraging on the more mature underwater remotely operated vehicle (UROV) technology, the ocean exploration has become more feasible and accessible. One important challenge is to realize reliable communication between the UROV and the control center on shore [1]. At present, there are various solutions, such as undersea cables transport, acoustic communication and underwater optical wireless communications (UOWC). The latter two wireless schemes provide great flexibility, whereas the acoustic communication exhibits long-range transmission capability, despite its much lower data rates. For UOWC, transmission data rate at Gigabit/s using visible light in green to blue region have been demonstrated [2]–[4], thus receiving much attention recently. Nevertheless, the practical realization of UOWC is challenging in the undersea environment. One of the challenging factors is the bubble-induced impediment to the UOWC link. Multiple causes, including atmospheric, benthic, and cavitation, would incur the formation of bubbles [5], as illustrated in Fig. 1. The primary sources of bubbles in the upper ocean are breaking waves, rains impacting on the sea surface, as well as man's activities such as that from ship propellers. For benthic sources of bubbles, the biological activities and the vents and seeps of gases escaping from the seafloor are possible to cause the generation of bubbles up to several millimeters in diameter. In [6], an experiment with emulated bubble effect has been conducted in a single-link UOWC system. It is demonstrated that bubbles can incur discontinuous blocking for the light signal and the corresponding UOWC transmission performance would be degraded seriously. Hence, the mitigation of bubble effect is an important issue for practical UOWC system. A statistical analysis of fading for UOWC in the presence of air bubble has been reported in [7]. However, both studies mainly focus on received optical signal power distribution and lack a comprehensive study on the link performance, such as bit error rate (BER) and outage probability.

Optical multiple input multiple output (MIMO) utilizing spatial multiplexing have been widely investigated to enhance the capacity of the VLC system [8]–[9], as well as in indoor visible light communication (VLC) systems with line-of-sight (LOS) drawback [10]. In [11], a precoding scheme is proposed under dynamic mobile receiver for an OFDM OWC system. An adaptive scheme is developed in [12], which can increase the overall performance of the VLC system, but also enhance the robustness of the system with a blocking by an object within the transmission area. However, the effect of bubbles in UOWC system is beyond a simple blocking or constant attenuation effect. The difference is that the bubbles are distributed discretely and dynamically in different sizes and the light can be reflected and refracted by the moving bubbles. The contribution of this paper is a comprehensive study of the optical transmission impairment by various bubble effects and the investigation of the robustness UOWC link by transmit and receive diversity.

In this paper, firstly, we experimentally investigate the bubble-induced impairment for UOWC. The link performance in terms of BERs under different bubble size and relative location with respect to the transmitter are investigated and analyzed. Secondly, we investigate the effectiveness of the spatial diversity using MIMO technique to increase the link robustness. With the help of a 2×2 MIMO system using repetition coding (RC) and maximum ratio combining (MRC), a 780-Mbit/s UOWC system is demonstrated with a link outage probability less than 1%, significantly decreased from 34.6%, for an optical link affected by bubbles.

Spatial diversity has been widely used in radio-frequency (RF) wireless transmission by utilizing multiple spatially distributed antennas to mitigate the Rayleigh fading in wireless channel [13]. There are different formats according to different system configurations: single input multiple output (SIMO), multiple input single output (MISO) and MIMO. On the one hand, to exploit the transmit diversity (MISO case), different space-time coding approaches can be employed, including repetition coding (RC) and Almouti-coded space-time block coding [14]–[16]. Here, we choose a low-complexity and effective scheme, the RC scheme, which can provide constructive power enhancement for intensity modulated/ direct detection (IM/DD) system.

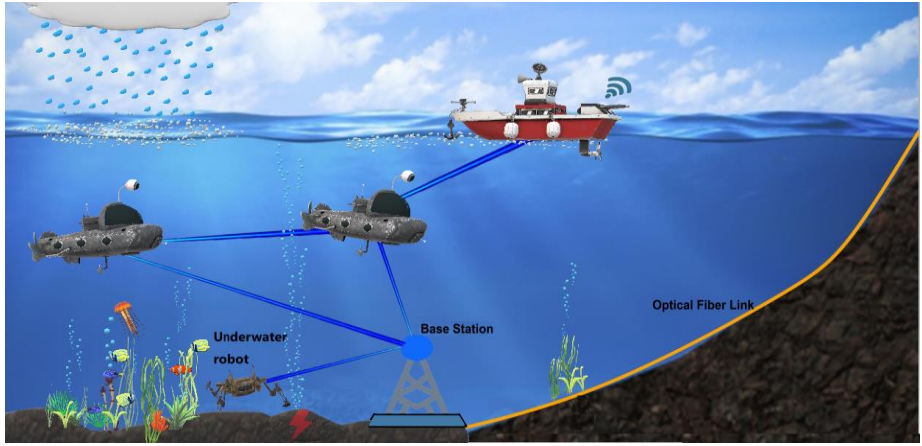

Fig. 1. Illustration of the ocean bubbles, UOWC links, and underwater communication networks

On the other hand, to exploit receive diversity (SIMO case), signals combination methods that include equal-gain combining (EGC), selective combining (SC), and maximum ratio combining (MRC) [17] are commonly used. The MRC is based on accurate channel state information at the receiver side, thus exhibiting the optimal signal-to-noise ratio (SNR) performance among the linear diversity combination approaches. It sums up the multiple received signals with optimal combining ratios that are proportional to the received signal strength, assuming the local noises are independent with equal power. The combined signal can be expressed as:

$$y = \sum_{i=1}^{N} h_i^* r_i = \sum_{i=1}^{N} h_i^* (h_i x + n_i) = \sum_{i=1}^{N} |h_i|^2 x + \sum_{i=1}^{N} h_i^* n_i \quad (1)$$

where $N$ is the total number of receivers and $h_i$ is the channel coefficient of $i$-th link. $r_i$ is the received signal of $i$-th receiver and it composes of the transmitted signal $x$ and noise term $n_i$. With MRC applied, the corresponding optimal SNR is given by:

$$SNR = \frac{E\left[(\sum_{i=1}^{N} |h_i|^2 x)^2\right]}{E\left[(\sum_{i=1}^{N} h_i^* n_i)^2\right]} = \frac{S \cdot \sum_{i=1}^{N} |h_i|^2}{N_0}. \quad (2)$$

Fig. 2 shows the experimental setup and the DSP block for investigating the UOWC performance in the presence of air bubbles. Two transmitters (TXs) and two receivers (RXs) are used in the experiment to realize various spatial diversity schemes. The space between two TXs and two RXs are 10 cm and 8 cm, respectively. For the two transmitter lanes, the original electrical signals are generated by an FPGA board equipped with a dual-channel digital-to-analog converter (DAC) operating at 1 GS/s sampling rate with a 16-bit acquisition accuracy. The FPGA board is used to interact with a personal computer (PC) in real-time to receive the digital signal from the PC and to control the DAC to generate the corresponding electrical signals. The data is first mapped to 16-quadrature amplitude modulation (16-QAM) signals, and then modulated as real-valued OFDM signals. The size of the inverse fast Fourier transform (IFFT) is 512, and out of which, 103 are modulated with data. This results in a signal bandwidth of 200 MHz, which is mainly limited by the PIN photodiode (PD). Taking the cyclic prefix (CP) length of 1/32 into consideration, the total data rate is 780 Mbit/s.

As the intensity of the generated signals is limited within 1 Vpp dynamic range, the drive strength of the signals is inadequate. Hence, two amplifiers are used to amplify the signals before injecting it into the transmitter-side light sources. The light sources used in the experiment are blue laser-diodes (LDs) (OSRAM PL450), with a center wavelength of 450 nm, within the open spectrum window for underwater transmission. Two bias-tees are used to combine a DC component and the amplified signals to drive the LDs. The amplification gain and the bias voltage are optimized at 9 dB and 5.4 V, respectively. Note that, the maximum power of the LDs is 80 mW in the experiment. A water tank of size 60 cm × 30 cm × 40 cm is used in the experiment for emulating the underwater channel. Bubbles are added in the channel by an air pump connected to different bubble dispensers for generating different size of bubbles. The total transmission link is short, ~80 cm, as the study is mainly focusing on the bubble effects and no lens is used.

At the receiver side, two PDs (Hamutuas S10784 with a cutoff frequency of 250 MHz) equipped with trans-impedance amplifiers (TIAs) are used to realize an optical-to-electrical conversion to recover the received optical signals. The photosensitive area of the PD has a diameter of 3.0 mm. A dual-channel analog-to-digital converter (ADC) carried by the same FPGA, receives the two-lane electrical signals. After uploading the captured data into the PC, an analysis is performed to evaluate the transmission performance under the bubble effect with and without spatial diversity.

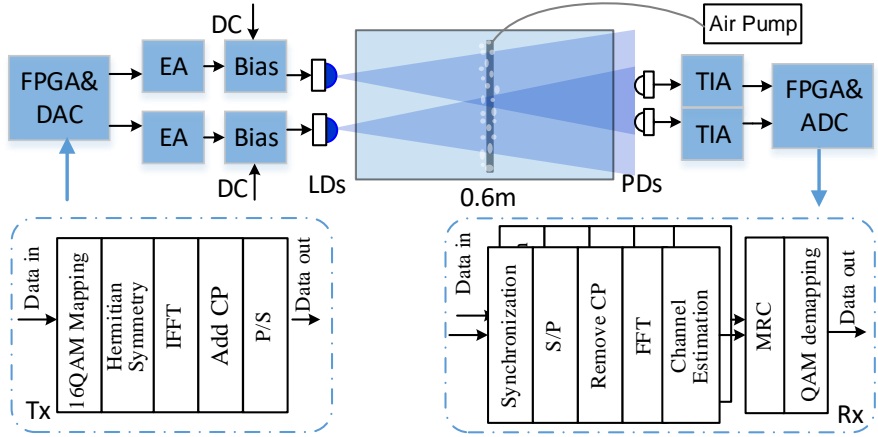

Fig. 2. Experimental setup and DSP block.

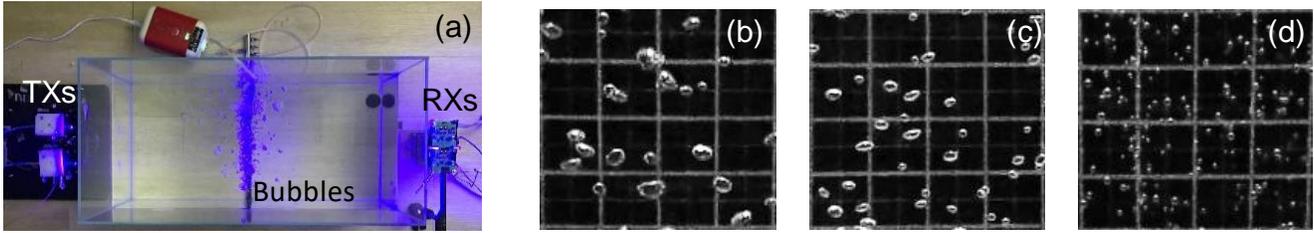

Fig. 3. (a) The experimental testbed. Images of bubbles captured in the experiment with an estimated average size of (b) Large size: ~ 15 mm², (c) Medium size: ~ 6 mm², (d) Small size: ~2 mm². Note that each square in the background is 10 mm × 10 mm.

In addition to the investigation of different spatial schemes, the transmission performance versus bubbles size and positions is also investigated experimentally. To study the effect of air bubble size, we use bubble dispenser with different hole size on it, while the flow rate of the air pump is fixed at 2 liters per minute (l/m) throughout the experiment. In this way, with the same total air volume used, different bubble clusters can be obtained. Three sizes of bubble are generated and investigated in this experiment, as shown in Fig. 3(b)–3(d). As estimated from the captured photos, they are categorized as large-, medium-, and small-size bubbles, with an approximate average size of 15 mm², 6 mm², and 2 mm², respectively. Meanwhile, we put the bubble tube at three different positions in the water tank to further investigate the system performance. The three positions are (i) near the transmitter (P1: 10cm), (ii) approximately in the middle of the transmitter and the receiver (P2: 40cm), and (iii) near the receiver (P3: 70 cm).

We first investigate the SNR and BER performance without bubble, which is to be used as a reference for further evaluation of the link performance. Four schemes are used for comparison, i.e. SISO (1×1), SIMO (1×2), MISO (2×1), and MIMO (2×2). The received optical power for the two receivers are -3.2 dBm and -2.8 dBm (with one LD), respectively, For the two-LD case, the received power are increased to -0.6 dBm and -0.4 dBm, repsectivley. Fig. 4 (a) shows the SNR of the four schemes. The power fading is due to the limited bandwidth of the PIN PD. When the optical spot at the receiver is much larger than the receiver area, no matter whether two LDs (2×1) or two PDs (1×2) are used, the received optical power will be nearly doubled when compared to the SISO case, resulting in an average 2.5-dB SNR improvement. Fig. 4 (b) and (c) show the BER versus the bandwidth used and the BER distribution at a bandwidth of 200 MHz, respectively. For the four schemes, the average BER are $1.31×10^{-2}$, $5.58×10^{-3}$, $3.92×10^{-3}$, and $1.42×10^{-3}$, respectively.

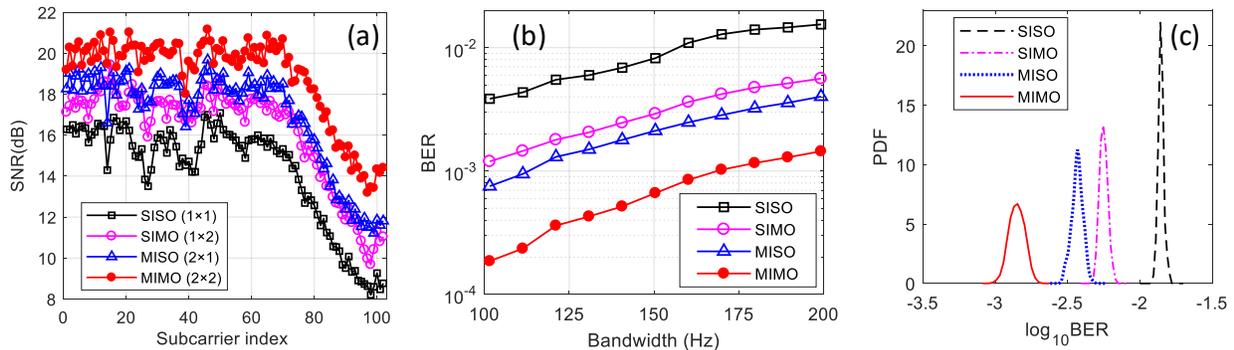

Fig. 4. The performance of UOWC system without bubble, (a) SNR versus subcarrier index with a bandwidth of 200 MHz, (b) BER versus bandwidth used, (c) BER distribution at a bandwidth of 200 MHz.

We then experimentally investigate the BER performance under the effect of bubbles with three different bubble sizes and at three different positions. For each scheme, 500 packets are captured to analyze the bubble-induced impairment and the mitigation capability provided by spatial diversity. 200 samples are shown in each figure of Fig. 5 to illustrate the bubble-induced effect under different scenarios to capture the variations of BER due to the dynamic bubbles. For the SISO scheme, it is clear that the bubble may block the line-of-sight light path and cause severe performance degradation. As shown in Fig. 3(b)–3(c), when the bubble size decreases, the number of bubble increases. When the bubble size is smaller, the blocking effect is less significant, which may give rise to partial light detection at the receiver. Therefore, as shown in Fig. 5(b), 5(e), 5(h), BER distribution undergoes from less frequent but larger variation to more frequent but smaller variation. Compared with the bubble size, the position of bubbles results in more distinct performance. With the bubble position moving away from the transmitter, the number of packets suffering severe fading subsides and the BERs of other packets have a larger variance. This is due to the light reflection and scattering with the bubbles. Similar performance is observed for both MISO and SIMO schemes. By providing a diversity gain of two, the probability of deep fading decreases significantly compared with the SISO scheme.

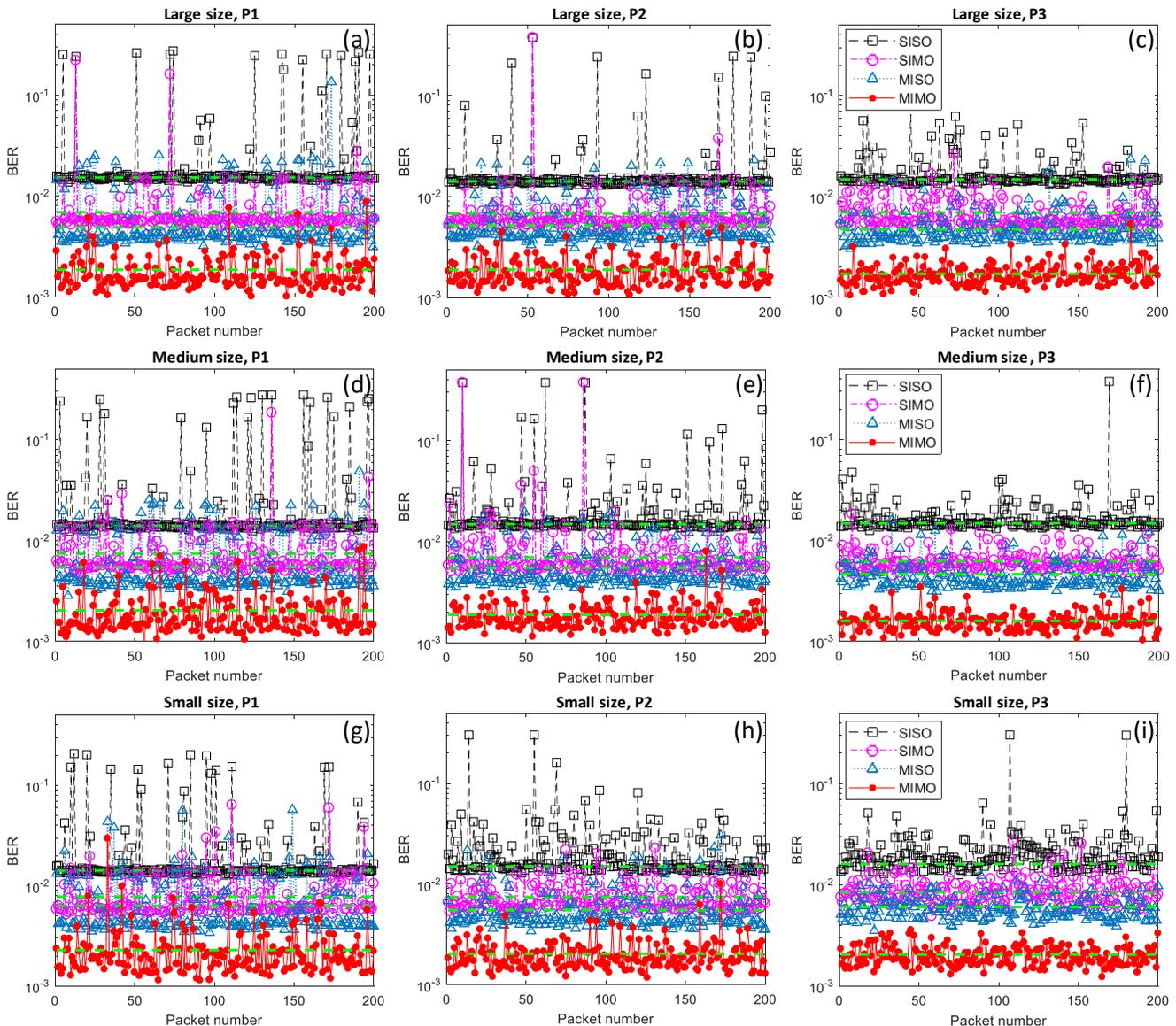

Fig. 5. The BER performance of 200 packets with (a)–(c) large-size bubbles, (d)–(f) medium-size bubbles, and (g)–(i) small-size bubbles, at a distance from TXs of P1: 10 cm, P2: 40 cm, and P3: 70 cm, respectively.

The MIMO case (the red cruve in Fig. 5), which further doubles the diversity gain, provides the most robust link as well as the lowest average BER. The performance is relatively constant with different bubble sizes at the same location. Next, we consider the effect of bubble location. The BERs show a decreased variance when the bubble location moves away from the transmitters. This also indicates a consistent trend that the complete blocking for each SISO link becomes less probable and scattered light increases the possibility of receiving partial signal. The green dash lines show the average BER of the packets that are below soft-decision forward error correction (SD-FEC) limit ($2\times10^{-2}$), which remains relatively constant for all the cases.

We then analyze the probability density function (PDF) of the BERs of 500 packets, as shown in Fig. 6. For all cases, the BER distribution spreads wider, compared with the cases without bubbles (Fig. 4(c)). The SISO case shows the largest difference in PDF under various scenarios. For the small bubble case at position 3 (near receiver), the peak of the PDF distribution for SISO reduces significantly, indicating the BER has a wider spread. In general, by exploiting the spatial diversity, the distribution for almost all scenarios remains relatively constant, which further verifies the observation from Fig. 5.

At last, we analyze the BER variance, packet loss rate (PLR) and the average BER of success received packets to further study the bubble-induced effect and to verify the diversity gain provided by different schemes. The BER variance in Table 1 further confirms the previous observation that the vairance is larger when bubbles are closer to the transmitter. The PLRs are given in Table 2. Note that, for the loss packet number less than five, the PLR is denoted as <1%. For the SISO case, as the distance increases or the bubble size decreases, the PLR shows a rising trend, whereas for the SIMO and MISO cases, the PLR decreases. This is because the original performance of SISO is near the FEC limit and more frequent flcution in BER leads to more packets having BER above FEC limits. For MIMO scheme, the outage probability is <1% for all the scenarios. By providing multiple independent light paths in the UOWC link, MIMO scheme significantly reduces the BER fluctuation and increases the link robustness. The average BER is shown in Fig. 7. The performance under small bubble case exhibits the worst BER.

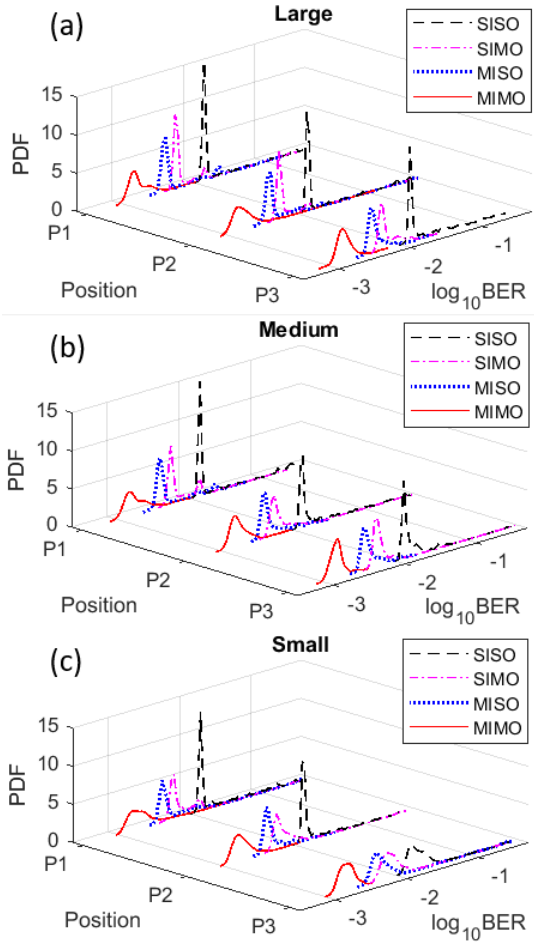

Fig. 6. The PDFs of the BER performance under scenarios with different bubble sizes, (a) large, (b) medium, and (c) small.

Table 1. BER variance

| Size/position | | SISO | SIMO | MISO | MIMO |
|---|---|---|---|---|---|
| L | P1 | 2.83E-03 | 2.80E-04 | 2.09E-03 | 9.65E-07 |
| | P2 | 2.63E-03 | 1.08E-04 | 6.46E-04 | 5.71E-07 |
| | P3 | 8.55E-04 | 8.04E-06 | 5.61E-06 | 2.23E-07 |
| M | P1 | 3.81E-03 | 1.42E-04 | 3.83E-04 | 1.33E-06 |
| | P2 | 2.90E-03 | 8.14E-04 | 1.15E-05 | 4.49E-07 |
| | P3 | 5.59E-04 | 5.45E-06 | 3.37E-06 | 1.46E-07 |
| S | P1 | 2.25E-03 | 7.26E-04 | 3.03E-04 | 2.41E-06 |
| | P2 | 1.40E-03 | 1.03E-04 | 1.22E-05 | 7.15E-07 |
| | P3 | 7.02E-04 | 8.93E-05 | 2.63E-06 | 1.63E-07 |

Table 2. Packet loss rate

| Size/position | | SISO | SIMO | MISO | MIMO |
|---|---|---|---|---|---|
| L | P1 | 13.0% | 3.4% | 9.6% | <1% |
| | P2 | 14.6% | 1.6% | 7.8% | <1% |
| | P3 | 20.4% | <1% | <1% | <1% |
| M | P1 | 16.4% | 4.4% | 5.6% | <1% |
| | P2 | 20.6% | 5.2% | <1% | <1% |
| | P3 | 21.0% | <1% | <1% | <1% |
| S | P1 | 24.4% | 3.2% | 6.4% | <1% |
| | P2 | 17.0% | 2.8% | 2.0% | <1% |
| | P3 | 34.6% | 2.4% | 1.2% | <1% |

In summary, we experimentally investigate the performance under different bubble sizes and positions. The statistics of the PLR and BER through a UOWC channels with bubbles are analyzed. For SISO scheme, the results show that as the distance to the transmitter side increases or the bubble size decreases, the less possibility of deep fading occurs. Without considering the loss packet, the average BER remains similar. Moreover, via spatial diversity, including SIMO, MISO, and MIMO schemes, it significantly reduces the BER fluctuation and increases the link stability. A robust 780-Mbit/s 2×2 MIMO system is achieved under different scenarios of bubble size and position. Enhancement in both BER (reduced from $1.31\times10^{-2}$ to $1.42\times10^{-3}$) and link reliability (outage probability reduced from the worst case 34.6% to less than 1% for all cases) are achieved.

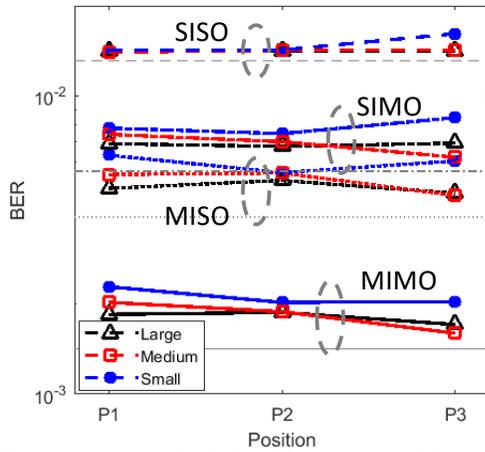

Fig. 7. The average BER of packets below SD-FEC. (The grey lines show the reference BER of each scheme without bubbles.)

This work was supported in part by HKSAR UGC/RGC grants (GRF 14215416 and GRF 14201217).